\documentclass[12pt]{article}

\usepackage{a4}
\usepackage{a4wide}

\usepackage{amsmath}
\usepackage{amscd}
\usepackage{amsfonts}
\usepackage{amssymb}
\usepackage{mathrsfs}
\usepackage{fancybox} 
\usepackage{enumerate}
\usepackage{accents} 

\usepackage{bm}
\usepackage{amscd}
\usepackage{graphicx}
\usepackage[dvips]{color}
\usepackage{epsfig}

\makeatletter

\@addtoreset{equation}{section}
\makeatother

\usepackage{amsthm} 
\usepackage{ascmac}  

\def\={\stackrel{\bullet}{=}}

\def\({\left(}
\def\){\right)}
\def\[{\left[}
\def\]{\right]}

\def \be {\begin{equation}}
\def \ee {\end{equation}}
\def \beqa {\begin{eqnarray}}
\def \eeqa {\end{eqnarray}}
\def \beal#1 {\begin{align}#1\end{align}}
\def \bes#1 {\begin{equation}\begin{split}#1\end{split}\end{equation}}
\def \nn {\notag\\}

\def\ket#1{|#1 \rangle}

\def\aver#1{\left\langle #1 \right\rangle}

\makeatletter

\@addtoreset{equation}{section}
\makeatother

\begin{document}

\begin{titlepage}
\title{
\begin{flushright}
\normalsize{ 
YITP-17-72}
\end{flushright}
       \vspace{1.5cm}
Flow equation, conformal symmetry and AdS geometry
\vspace{1.5cm}
}
\author{
Sinya Aoki\thanks{saoki[at]yukawa.kyoto-u.ac.jp
},\; Shuichi Yokoyama\thanks{shuichi.yokoyama[at]yukawa.kyoto-u.ac.jp
}
\\[25pt] 
{\it Center for Gravitational Physics,} \\
{\it Yukawa Institute for Theoretical Physics, Kyoto University,}\\
{\it Kitashirakawa-Oiwakecho, Sakyo-Ku, Kyoto, Japan}
\\[10pt]
}

\date{}
\maketitle

\thispagestyle{empty}


\begin{abstract}
\vspace{0.3cm}
\normalsize
We argue that the Anti-de-Sitter (AdS) geometry in $d+1$ dimensions naturally emerges from an arbitrary conformal field theory in $d$ dimensions using the free flow equation.
We first show that an induced metric defined from the flowed field generally
corresponds to the quantum information metric, called the Bures or Helstrom metric, if the flowed field is normalized appropriately.
We next verify that the induced metric computed explicitly with the free flow equation always becomes the AdS metric when the theory is conformal.
We finally prove that the conformal symmetry in $d$ dimensions converts to the AdS isometry in $d+1$ dimensions after $d$ dimensional quantum averaging.
This guarantees the emergence of AdS geometry without explicit calculation.
\end{abstract}
\end{titlepage}

\section{Introduction} 
The AdS/CFT (or Gravity/Gauge theory) correspondence \cite{Maldacena:1997re} 
is a promising tool to crack a hard problem in strongly coupled gauge theories (see \cite{Aharony:1999ti,Klebanov:2000me,DHoker:2002nbb} for some reviews), but 
is still mysterious even after many pieces of evidence and application appeared after the first proposal.
Although the correspondence may be a manifestation of the closed string/open string duality, an alternative understanding might exist due to its holographic property.

One important mystery is the precise mechanism how the AdS radial direction emerges from conformal field theory (CFT). It may be a common sense that the AdS radial direction is emergent as a renormalization scale of dual CFT as both behave similarly  under the dilatational symmetry \cite{Maldacena:1997re}, and 
this viewpoint worked out for renormalization group flows triggered by relevant deformations \cite{Girardello:1998pd,Distler:1998gb}.
However a direct approach to search the Wilsonian cutoff corresponding to the sharp cutoff on AdS radial direction \cite{Heemskerk:2010hk} is still far from a clear answer, since the ordinary Wilsonian renormalization gives rise to non-local interaction in the bulk interpretation. See also \cite{Lee:2010ub}.%
\footnote{This situation may drastically change in the AdS$_3$/CFT$_2$ correspondence. A clear dictionary was proposed in \cite{McGough:2016lol}.}

While several approaches were developed to {\it grab the tail} of AdS radial direction and construct bulk dynamics from CFT  \cite{Banks:1998dd,Balasubramanian:1999ri,Bena:1999jv,Lifschytz:2000bj} (see also \cite{VanRaamsdonk:2009ar,Nozaki:2012zj}),
one of the present authors, together with his collaborators, has proposed an alternative method to define a geometry from a quantum field theory and explicitly calculated the metric from several quantum field theories  \cite{Aoki:2015dla,Aoki:2016ohw,Aoki:2016env}.
In Ref.~\cite{Aoki:2015dla}, the method was proposed and applied to the O$(n)$ non-linear sigma model in 2-dimensions, and  the 3 dimensional metric in the large $n$ limit was shown to describe an AdS space in the massless limit.
In Ref.~\cite{Aoki:2016ohw},
it was shown in the large $n$ limit that 
the induced metric describes an AdS$_{d+1}$ space with $d\ge 3$ in the UV limit and with $d \ge 1$ in the IR limit, if  the method is applied to the massive O$(n)$ $\varphi^4$ model and
an appropriate normalization is introduced for the flowed field.
In Ref.~\cite{Aoki:2016env}, the large $n$ expansion was performed for the massless O$(n)$
$\varphi^4$ model in 3 dimensions. While the induced metric describes the AdS$_4$ space at the leading order, the next leading order corrections make the space asymptotically AdS only in UV and IR limits with different radii.
 
By observing that the induced metric always gives the AdS metric when the theory is conformal, it is natural to expect that symmetry plays a key role behind these results. In this paper, we investigate a direct relation between CFT and the AdS metric in this framework. The goal of this paper is to generalize the previous results to an arbitrary conformal field theory incorporating symmetry argument.  

The rest of this paper is organized as follows. 
In Sec.~\ref{sec:Metric}, after brief explanation of the proposal of Ref.~\cite{Aoki:2015dla},
we first show that our metric corresponds to the information metric stressing importance of the field normalization introduced in Ref.~\cite{Aoki:2016ohw}.
In Sec.~\ref{sec:Isometry},  we explicitly derive the AdS metric directly from the CFT.
We then prove that the induced metric in $d+1$ dimension possesses the isometry of the AdS space as a consequence of the conformal symmetry in $d$ dimensions.
Sec.~\ref{sec:Discussion} is devoted to summary and discussion.

\section{Gradient flow and information metric}
\label{sec:Metric}
\noindent
In this section, we briefly review the proposal in Ref.~\cite{Aoki:2015dla,Aoki:2016ohw} to define a $d+1$ dimensional induced metric from a $d$ dimensional quantum field theory.
We also show that the metric defined in this way with appropriate normalization can be interpreted as a quantum information metric, 
called the Bures or Helstrom metric. \\

\subsection{Gradient flow and induced metric}
We consider an $n$ real component scalar field $\varphi(x)$ in $d$ dimensions, whose quantum dynamics is controlled by the action functional $S(\varphi)$.  The flowed field $\phi$ is defined from $\varphi$ with the initial condition $\phi(x;0)=\varphi(x)$  through the flow equation as  
\beqa
\frac{\partial \phi^a(x;t)}{\partial {t}} = - \left.\frac{\delta S_f(\varphi)}{\delta \varphi^a(x)}\right\vert_{\varphi(x)\to\phi(x;t)},
\eeqa
where 
the flow time $t$ {has the (length)$^2$ dimension}, 
$x$ is the $d$ dimensional coordinate system, $a=1,2,\cdots, n$ labels a component of the scalar field, and $S_f(\varphi)$ is an appropriate action for $\varphi$, which is not necessarily related to the original action $S(\varphi)$ in general. Particularly when they coincide, the flow is called the gradient flow \cite{Narayanan:2006rf,Luscher:2010iy, Luscher:2009eq,Luscher:2013vga}.  In the case of the free flow ({\it i.e.} $S_f$ is the free action), the flow equation becomes the heat equation. Thus the flow equation defines a procedure to smear the original field $\varphi$ into a smeared field $\phi$,  correlation functions of which are all finite at $t > 0$.

A $d+1$ dimension metric operator is given by
\beqa
\hat g_{MN} (x;t) := R^2 \sum_{a=1}^n {\partial \sigma^a(x;t) \over \partial z^M} {\partial \sigma^a(x;t) \over \partial z^N}, 
\eeqa
where  {$R$ is a constant with  the length dimension,  and}
$z^M=(x^\mu, \tau)$ with {$\tau=\sqrt{2 d t}$}, which is regarded as the $d+1$ dimensional coordinates after $d$ dimensional quantum averaging, and $\sigma^a(x;t)$ is the {(dimensionless)} normalized flowed field defined as
\beqa
\sigma^a(x;t) := \frac{\phi^a(x;t)}{\sqrt{\langle \sum_{a=1}^n \phi^a (x;t)^2 \rangle_S}} .
\eeqa
Here $\langle O(\varphi) \rangle_S$ denotes the quantum average with the $d$ dimensional action $S$, given by 
\beqa
\langle O(\varphi) \rangle_S := \frac{1}{Z} \int {\cal D}\varphi\, O(\varphi)\, e^{-S(\varphi)}, \
Z = \int {\cal D}\varphi\,  e^{-S(\varphi)}.
\eeqa
Since $\sigma^a(x;t)$ always satisfies $\displaystyle\sum_{a=1}^n \left\langle \sigma^a(x;t) \sigma^a(x;t) \right\rangle_S =1$, we call this normalization condition the non-linear sigma model (NLSM) normalization, whose importance becomes clear in the next subsection.
The vacuum expectation value {(VEV)} of the metric is denoted as  $g_{MN}(z) := \langle \hat g_{MN}(x;t) \rangle_S$, {whose fluctuations are suppressed in the large $n$ limit.}
(See Ref.~\cite{Aoki:2015dla,Aoki:2016ohw,Aoki:2016env} for more details.)\\ 

\subsection{Information metric}
\label{InformationMetric}
In this subsection, we show that $g_{MN}(z)$ defined in the previous subsection is equivalent to a quantum information metric, called the Bures (or Helstrom) metric. The NLSM normalization is important to show this.

The Bures metric for the density matrix is defined from the infinitesimal distance between two density matrices $\rho$ and $\rho+d\rho$ as
\beqa
D(\rho,\rho+d\rho)^2 = \frac{1}{2}{\rm tr} (d\rho\, G),
\eeqa
where $G$ is the Hermitian 1-form operator implicitly given by
$
\rho\, G + G\, \rho = d\rho.
$
In particular for the density matrix $\rho$ of a pure state, the Hermitian operator is determined as $G=d\rho$ since $\rho^2=\rho$.

In order to apply to our case, we consider an eigenstate of the position operator as well as the flow time one denoted by $\ket{z}=\ket{(x, \tau)}$ and define the inner product of the state as
\beqa
\langle z \vert w \rangle := 
\sum_{a=1}^n \langle \sigma^a(x;t) \sigma^a(y;s)\rangle_S , \quad
w=(y,\sqrt{2d s}) .
\eeqa
Notice that the NLSM normalization guarantees $\langle z \vert z \rangle =1$.
Then the information metric for this pure state is computed as 
\beqa
R^2 D(\rho_z, \rho_{z+dz})^2 
&=&  g_{MN}(z) dz^M dz^N, 
\eeqa
where $\rho_z =\vert z\rangle \langle z\vert, $ and 
we used $\langle\sigma(x;t)\cdot \partial_M \sigma(x;t) \rangle_S =0$ and
\beqa
 \langle\sigma(x;t)\cdot \partial_M\partial_N  \sigma(x;t) \rangle_S
= -  \langle \partial_M \sigma(x;t) \cdot \partial_N  \sigma(x;t)\rangle_S  \notag 
\eeqa
with $A\cdot B := \sum_{a=1}^n A^a B^a$.
Our metric $g_{MN}(z)$ defines a distance in the space of the density matrices made of the pure states (in unit of $R$). 
In particular when the action $S$ of the original theory has O$(n)$ symmetry, such pure states defined above in an abstract way may be given by $\sum_{a}\sigma^a(x;t) \vert 0 \rangle$.%
\footnote{In the case of the $O(n)$ invariant mixed state $\rho_z :=\sum_a \sigma^a(x;t) \vert 0 \rangle \langle 0 \vert  \sigma^a(x;t)$, we have $G = n d\rho_z$ from  $\rho_z^2 =\rho_z/n$.}


\section{AdS isometry from conformal symmetry} 
\label{sec:Isometry}
In this section, 
we directly relate arbitrary conformal field theory in the flat $d$ dimensional space-time
with the AdS metric in $d+1$ dimensions.\footnote{The argument and calculation below hold just by changing the signature suitably when we consider the Euclidean flat space.}
We here assume that the CFT  contains a real scalar primary operator $\varphi(x)$ of a general conformal dimension $\Delta$ without specifying any concrete models for CFT.
{In this section, we are interested in the VEV of the metric operator, which determines the classical geometry induced from this  primary operator,
though its ``quantum" fluctuations around the VEV exist for $n=1$ in this case.}   \\ 

\subsection{AdS metric from CFT}
We start with the free generally massive flow equation {with $n=1$}
\beqa
\frac{\partial \phi(x;t)}{\partial {t}} = {(\partial^2 - m^2)} \phi(x;t), 
\label{eq:free_flow}
\eeqa
where $\partial^2 = \eta^{\mu\nu} \partial_\mu\partial_\nu$. 
This is easily solved as
\beqa
\phi(x;t) = e^{ {t} {(\partial^2-m^2)}}\varphi(x) .
\eeqa

The two-point function of $\phi$ is evaluated as
\beqa
G_0(x;t|y;s)&:=& \langle \phi(x;t) \phi(y;s) \rangle_{\rm CFT} \nonumber \\
&=& {e^{-(t+s)m^2}} e^{({t}\partial^2_x + s\partial^2_y)} \langle\varphi(x)\varphi(y) \rangle_{\rm CFT},
\eeqa
where we used the subscript CFT instead of $S$, since the action is not specified.
The Poincare invariance and the scale transformation of $\varphi$ with $ \varphi(\lambda x) = \lambda^{-\Delta} \varphi(x)$
fix the form of $G_0$ such that
\beqa
G_0(x;t|y;s) = \frac{{e^{-(t+s)m^2}}}{{(t+s)^\Delta}} F_0\left(\frac{(x-y)^2}{{t+s}}\right)
\eeqa
where $F_0$ is a certain smooth function. 
Furthermore, the flow equation (\ref{eq:free_flow}) implies
\beqa
\frac{\partial}{\partial t}  G_0(x;t|y;s) =
{({\partial^2_x} -m^2)} G_0(x;t|y;s), 
\eeqa
which leads to $\Delta F_0(0) = -2 d F_0^\prime (0) $. 
Thus the two-point function of the normalized field $\sigma$ {becomes $m$ independent as}  
\beqa
G(x;t|y;s) &:=& \langle \sigma(x;t) \sigma(y;s) \rangle_{\rm CFT} \nonumber\\
&=&  \left(\frac{2\sqrt{ts}}{t+s}\right)^\Delta F\left(\frac{(x-y)^2}{{t+s}}\right),
\label{2ptsigma}
\eeqa
where $F(x) \equiv F_0(x)/F_0(0)$.
Hence $F(0)=1$ and $2 d F^\prime(0) = -\Delta$.\footnote{ 
Explicitly, $F(x)$ is computed as  
\beqa
F(x) = \frac{\Gamma(d/2)}{\Gamma(\Delta)\Gamma(d/2-\Delta)}&\displaystyle \int_0^1& dv\,  v^{\Delta-1}(1-v)^{d/2-\Delta-1}e^{-x v/4}. \nonumber 
\eeqa
}
Taking the $d+1$ dimensional coordinates as $z=(x,{\tau=\sqrt{2dt}})$,
the vacuum expectation value of the induced metric $\hat g_{MN}$ is 
\beqa 
g_{MN}(z) 
= R^2 \langle \partial_M\sigma(x;t) \partial_N\sigma(x;t) \rangle_{\rm CFT}. 
\eeqa
This is computed as ($g_{\mu\tau}(z)=g_{\tau \nu}(z)=0$)
\beqa
g_{\mu\nu}(z) &=&\eta_{\mu\nu} \frac{R^2 \Delta}{\tau^2}, \quad 
g_{\tau\tau}(z) =\frac{R^2 \Delta}{\tau^2},
\label{eq:AdS_metric}
\eeqa
which is nothing but the AdS metric with $R\sqrt\Delta$ its radius.
\\

\subsection{Isometry from conformal transformation}

In the previous subsection, 
we have explicitly calculated the induced metric $g_{MN}$ from CFT,
and have shown that it becomes  the AdS metric in the Poincare patch.
In this subsection, we argue the relation between the AdS metric and the CFT only from the symmetry.
Namely we show that the induced metric $g_{MN}(z) =\aver{\hat g_{MN}(x;t)}_{\rm CFT}$ possesses the isometry of the AdS space.  
This is necessary and sufficient since AdS is a maximally symmetric space, so that the metric is completely fixed by the isometry group $SO(2,d)$ up to an overall constant. 

We first relate the conformal transformation to the isometry of AdS.%
\footnote{ 
{This question was addressed in the early stage after the Maldacena's proposal in \cite{Jevicki:1998qs}, where the mechanism was studied how the special conformal transformation of the adjoint scalar fields in ${\cal N}=4$ super Yang-Mills theory was {\it metamorphosed} into the corresponding isometry transformation in the bulk.
We would like to thank T.~Yoneya for his valuable comment given in KEK Theory workshop 2017 ``East Asia Joint Workshop on Fields and Strings 2017''. }
}
The infinitesimal conformal transformation and the response of the primary scalar operator are given by 
\beqa
\delta x^\mu &=& a^\mu + \omega^\mu{}_\nu x^\nu + \lambda x^\mu + b^\mu x^2 -2 x^\mu (b_\nu x^\nu), \nonumber \\
\delta^{\rm conf} \varphi(x)  &=& - {\delta x}^\mu\partial_\mu \varphi(x)  -  {\Delta \over d}  (\partial_\mu \delta x^\mu) \varphi(x).
\label{conformalTr} 
\eeqa
Here $a^\mu, \omega^{\mu\nu}, \lambda, b^\mu$ are parameters of the transformation.  
Since the infinitesimal conformal transformation is quadratic in the coordinate $x$,  the normalized field $\sigma$, given  in terms of $\varphi$ by
\beqa
\sigma (x;t) = \frac{ \left(\sqrt{2t}\right)^{\Delta}}{\sqrt{F_0(0)}}\, e^{ {t \partial^2}}\, \varphi(x),
\eeqa
is transformed as%
\beqa
\delta^{\rm conf}\sigma(x;t)
&=&- \left\{{t} (\partial^2 \delta x^\mu) + 2 {t^2}  (\partial^\nu \partial^\rho \delta x^\mu) \partial_\nu \partial_\rho   + 2{t} (\partial^\nu \delta x^\mu) \partial_\nu + \delta x^\mu\right\} \partial_\mu \sigma(x;t) \nn &&
- {\Delta \over d}\left\{2 {t} (\partial^\nu \partial_\mu {\delta x}^\mu) \partial_\nu + (\partial_\mu {\delta x}^\mu) \right\} \sigma(x;t) . 
\label{conformalTrsigma}
\eeqa
Plugging \eqref{conformalTr} into \eqref{conformalTrsigma} we obtain 
\be
\delta^{\rm conf}\sigma(x;t)
= -\{ 2\lambda - 4 (b_\mu x^\mu) \}t\partial_t  \sigma(x;t)  - \{\delta x^\mu+2  {t} (d-2 - \Delta ) b^\mu  \}\partial_\mu \sigma(x;t) 
 + 4 {t}^2 b^\mu \partial_\mu \partial_t  \sigma (x;t), 
 \label{eq:conformalTrsigma2}
\ee
where we have used $\partial_t \sigma(x;t) =({\Delta \over 2t } + {\partial^2)} \sigma(x;t) $.
{Notice that a special conformal transformation of the normalized flow operator contains a higher derivative term as the last term in \eqref{eq:conformalTrsigma2}, which cannot be interpreted as an infinitesimal diffeomorphism in the bulk.}%
\footnote{ 
{Difficulty in the bulk interpretation of the special conformal transformation of a key operator was also observed in the bilocal field approach by using a vector model \cite{Das:2003vw}, where the special conformal transformation of the collective field mixes up fields with different spin. They speculated on the necessity of a suitable field redefinition to resolve this issue, though our answer may be different in the flow field approach as shown below. 
We would like to thank S.~Das for his valuable comment given in KIAS-YITP joint workshop 2017 ``Strings, Gravity and Cosmology".} 
}
{To deal with this we} rewrite eq.~(\ref{eq:conformalTrsigma2}) as
\beqa 
\delta^{{\rm conf}} \sigma (x;t) &=& \delta^{{\rm diff}} \sigma(x;t) + \delta^{{\rm extra}} \sigma (x;t), \\
\delta^{{\rm diff}} \sigma(x;t) &=& -(\bar\delta t \partial_t + \bar\delta x^\mu \partial_\mu)\sigma (x;t), \quad 
\delta^{{\rm extra}}\sigma(x;t) 
= 4 {t^2} b^\nu \partial_\nu (\partial_t + {\Delta +2 \over 2t })  \sigma (x;t),~~~~~~
\label{extraterm}
\eeqa
with $ \bar\delta x^\mu =\delta x^\mu + 2d { t}  b^\mu,   
\bar\delta t =(2\lambda - 4 (b_\mu x^\mu) )t$. 
By setting $\tau^2 =2 d {t}$, the transformation $\delta^{\rm diff}$ can be rewritten as 
\beqa
\bar\delta x^\mu =\delta x^\mu + \tau^2 b^\mu   , \quad 
\bar\delta \tau  = (\lambda - 2  (b_\mu x^\mu))\tau .
\eeqa
This is nothing but the isometry transformation of the AdS space whose metric is given by $ds^2_{{\rm AdS}} \propto \displaystyle { d\tau^2 + dx^\mu{}^2 \over \tau^2} .$ 

The conformal transformation of the induced metric operator is computed as 
\be
\delta^{\rm conf}  \hat g_{MN}(x;t) 
=\delta^{\rm diff}  \hat g_{MN}(x;t) +R^2 \lim_{(y;s)\to (x;t)} {\partial \over \partial z^M} {\partial \over \partial w^N}\left\{ \delta^{{\rm extra}}\sigma(x;t) \sigma(y;s)  +\sigma(x;t) \delta^{{\rm extra}}\sigma(y;s)\right\}. 
\label{eq:confTrMetric}
\ee
The first term is 
\beqa 
\delta^{\rm diff} \hat g_{MN}(x;t) = - \bar\delta z^K \partial_K \hat g_{MN}(x;t) - \partial_M \bar\delta z^K \hat g_{KN}(x;t)  - \partial_N \bar\delta z^K \hat g_{MK}(x;t), 
\eeqa
which is nothing but the diffeomorphism of the metric tensor in the $d+1$ dimensions.

Thus our task is to show that the second term in eq.~(\ref{eq:confTrMetric}) vanishes in the vacuum expectation value.  
By using \eqref{2ptsigma} and \eqref{extraterm} the term is computed as 
\be
\aver{\delta^{{\rm extra}}\sigma(x;t) \sigma(y;s)  + \sigma(x;t)\delta^{{\rm extra}}\sigma(y;s)  } 
= -8 { (\sqrt{4 t s})^{\Delta} \over  (t+s)^{\Delta+2}}(t  -s) {b_\mu(x-y)^\mu }{(x-y)^2}
 F''\left({(x-y)^2 \over { t+s}}\right).   
\ee
Then it is easy to see
\beqa
 \lim_{(y;s)\to (x;t)} {\partial \over \partial z^M} {\partial \over \partial w^N}\aver{\delta^{{\rm extra}}\sigma(x;t) \sigma(y;s)  + \sigma(x;t)\delta^{{\rm extra}}\sigma(y;s)  } = 0.
\eeqa
{We stress that this happens only when the conformal transformation is decomposed as \eqref{extraterm}.}
Note that the quantum averaging and the differentiation commute since all correlation functions of $\sigma$ are finite as long as the flow time is non-zero.
Therefore we obtain 
\be 
\aver{\delta^{\rm conf} \hat g_{MN}(x;t)} =\aver{\delta^{\rm diff} \hat g_{MN}(x;t)}.  
\ee

Since the conformal invariance of the two point function of the primary scalar operator implies that $\aver{\delta^{\rm conf} \hat g_{MN}(x;t)} =0$, it follows that $\aver{\delta^{\rm diff} \hat g_{MN}(x;t)}=0$. This means that the induced metric $g_{MN}(z)$ satisfies the Killing equation of the AdS space as
\beqa
\delta^{\rm diff} g_{MN}(z) = - \bar\delta z^K \partial_K  g_{MN}(z) - \partial_M \bar\delta z^K  g_{KN}(z)  - \partial_N \bar\delta z^K  g_{MK}(z)  = 0, 
\eeqa
which implies that the induced metric must be the AdS metric up to an overall constant. 
This completes the proof of our claim. 
\\

\section{Discussion} 
\label{sec:Discussion}
In this paper an induced geometry by a flow equation from a quantum field theory was investigated. 
The induced metric was shown to appear as a quantum information metric, which measures a distance in the space of the pure states constructed by scalar fields in a general quantum field theory. 
In a conformally symmetric situation, it was shown that the induced metric matches the AdS one when the flow equation is free.%
\footnote{{
This result does not depend on mass $m$  in the flow equation, which violates the conformal symmetry,
thanks to the NLSM normalization.}}
This agreement was confirmed only by using symmetry without any explicit computations of the metric.
{An appearance of the AdS metric from CFT in our method relies on the following two facts.
(a) The field $\sigma$ used to define the metric operator is dimensionless thanks to the NLSM normalization. If $\phi$ were used instead, one would not obtain the AdS metric.   
(b) The VEV of the metric operator is UV finite thanks to the free flow equation.
If the VEV were UV divergent, one would not  obtain the AdS metric due to additional dimensionful quantities introduced through renormalization.   
}

So far any relation between the induced metric formalism presented in this paper and other approaches to see dual geometry mentioned in the introduction is not known. 
It may be reasonable to think that there is no relation between them since, for example, the procedure to renormalize fields in quantum field theory and that to smear operators are generally independent. 
Still we expect that the results and technique developed in this paper, particularly the symmetry argument, will become useful to study the AdS geometry from CFT by {other methods}. 
{For example, it may be possible to define an induced metric similarly in the Wilsonian renormalization approach to the dual gravity. Then it would be interesting to see whether the metric becomes the AdS one or not.}  
 
In this paper we assume that the background of quantum field theory or conformal field theory is flat. 
It would be interesting to extend the presented calculation to curved backgrounds. 
In particular  it would be curious to check whether the induced metric from CFT on a curved space-time is still of the AdS form in a different coordinate system from the Poincare patch. 

A challenging but important issue is whether this formalism encodes the gravitational dynamics or not. 
The first step toward this goal may be to see how a linearized Einstein gravity is encoded in this formulation, as shown in a different method to derive dual bulk dynamics by using the entanglement entropy \cite{Faulkner:2013ica,Bhattacharya:2013bna}.
For this analysis it will be necessary to specify a concrete model to test the proposal such as an O$(n)$ sigma model, since bulk dynamics is dependent on each CFT. {Note that the $1/n$ expansion becomes important to see the dual bulk dynamics beyond the geometry in AdS/CFT correspondence. }
A virtue of this formulation is that observables in this formalism are correlation functions of scalar fields in quantum field theory, which admit analytic computation by the ordinary technique of the $1/n$ expansion  \cite{Aoki:2016env}, so that one can proceed by checking one's guesswork explicitly by hand.  

We hope to report on resolutions on these issues in the near future. \\

\section*{Acknowledgment}

S. A. would like to thank Dr. G.~Ishiki for his comment on the information metric.
{We would also like to thank S.~Das and T.~Yoneya for valuable comments.} 
S. A. is supported in part by the Grant-in-Aid of the Japanese Ministry of Education, Sciences and Technology, Sports and Culture (MEXT) for Scientific Research (No. JP16H03978),  
by a priority issue (Elucidation of the fundamental laws and evolution of the universe) to be tackled by using Post ``K" Computer, 
and by Joint Institute for Computational Fundamental Science (JICFuS).

\bibliographystyle{utphys}
\bibliography{CFT2AdS}

\end{document}